\documentclass[11pt,prd,preprintnumbers,eqsecnum,superscriptaddress,nofootinbib]{revtex4}
\usepackage[usenames,dvipsnames]{xcolor}
\usepackage{graphics}
\usepackage{graphicx}
\def\be{\begin{equation}}
	\def\ee{\end{equation}}
\def\ba{\begin{array}}
	\def\ea{\end{array}}
\def\bea{\begin{eqnarray}}
	\def\eea{\end{eqnarray}}

\begin{document}
	
	\begin{center}
		{\Large \bf Electromagnetic $N+\gamma^{\ast}\longrightarrow N^{*}$ transition form factors of nucleons from the hard-wall AdS/QCD model }
		
		\vskip 1. cm
		{Shahin Mamedov $^{a,c,e,f}$\footnote{corresponding author: sh.mamedov62@gmail.com }} and
		{Shahnaz Taghiyeva $^{b,d,e}$\footnote{shahnazilgarzade@gmail.com}},
		
		\vskip 0.5cm
		
		{\it $^a\,$ Institute for Physical Problems, Baku State University,
			
			Z. Khalilov street 23, Baku, AZ-1148, Azerbaijan,}
		\\ \it \indent $^b$Theoretical Physics Department, Physics Faculty, Baku State University,
		
		Z.Khalilov street 23, Baku, AZ-1148, Azerbaijan,
		\\ \it \indent $^c$Institute of Physics, Ministry of Science and Education,
				
		H. Javid avenue 33, Baku, AZ-1143,
		Azerbaijan,\\ \it \indent $^d$Shamakhy Astrophysical Observatory, Ministry of Science and Education,
		
		H.Javid 115, Baku, AZ 1000, Azerbaijan,		
		\\ \it \indent $^e$Center for Theoretical Physics, Khazar Univertsity,
		
		41 Mehseti Street, Baku, AZ-1096, Azerbaijan,
		\\  \it \indent $^f$ Lab. for Theor. Cosmology, International Centre of Gravity and Cosmos, Tomsk State University of Control Systems and Radio Electronics (TUSUR),
  
  634050 Tomsk, Russia. \\

	\end{center}
	
	\centerline{\bf Abstract} \vskip 4mm

The electromagnetic transition form factors for the $N+\gamma^{\ast}\longrightarrow N^{*}$ transition between the ground and excited states of nucleons is studied in the framework of the hard-wall model of AdS/QCD. The 5 dimensional equation of motion was solved for the fermion and vector fields. The profile function of the spinor field and bulk-to-boundary propagator of the vector field are presented. The interaction Lagrangian includes other kinds of terms in addition to the minimal coupling term. Using the AdS/CFT correspondence between the generating functions in the bulk and boundary theories, an expression for the transition form factors is obtained from the bulk action for the interaction between the photon and nucleon fields. We consider the $N^{*}(1440,1535,1710)\rightarrow N$ transitions and plot the Dirac, Pauli and electric, magnetic form factors dependencies on momentum transfer. Also, plots for the helicity amplitudes have been presented and compared to experimental data. The transition radii obtained within the soft-wall model are close to the experimental data for the radii of the nucleons at ground states.

	\vspace{1cm}
	
	\section{Introduction}
	AdS/CFT correspondence, also known as gauge-gravity duality
	is one of the most exciting discoveries in modern theoretical
	physics. This correspondence was adapted to describe the low-energy
	dynamics of QCD in the simplest models known as AdS/QCD models.
	These models are useful for understanding properties of QCD at low
	energies such as chiral symmetry breaking, hidden local symmetry,
	vector meson dominance at large $N$ and high energy regimes. There
	are two approaches in AdS/QCD correspondence called as top-down
	and bottom-up approaches [1]. The bottom-up approach describes
	various low-energy phenomenology of QCD. The hard- and soft-wall models are
	the typical ones in the bottom-up approach. Mass spectra, decay
	and coupling constants, form factors were calculated in the
	framework of these models and gives the results in a good agreement with the data [2-6]. Holographic QCD models are useful  as
	well as for the study of internal structure of hadrons, including
	their excitations by means of study of the transition form factors. 
The last decade has been observed a progress in the experimental and theoretical studies of  nucleon resonances (the  radial/orbital nucleon excitations with $J^{P}=1/2^{\pm}$,  $J^{P}=3/2^{\pm}$). Electro- and photoproduction of the resonance states helps us to study their internal structure, wave functions and interactions between constituents. The nucleon resonance $ R(1440)$ with  $J^{P}=1/2^{+}$  (the Roper resonance $P_{11}$ or simply $R$)  is the lowest-lying excitation of the nucleon and its structure issue is a longstanding question.  Studies in the  constituent quark model (CQM) framework  showed that the inner structure of the Roper is possibly more complicated than the structure of the other lightest baryons. It was found that the observed mass of the Roper resonance is low and the decay width is large in comparison with the values predicted   by the CQM  [7]. The Roper state is  the first $(2S)$ radial excitation of the nucleon ground state $s^{3} [3]_{X}$. But this description fails to explain either the large decay width $\Gamma$ R $\simeq$ 300 $MeV$ or the branching ratios for the  $\pi N$  $(55\%-75\%)$  and $\sigma N$ $(5\% -20\% )$ decay channels [7-9].

 The elementary theory of strong interactions QCD provides a framework, which is directly usable only at high momentum transfers. Nevertheless, the discussed data  [7-11] span the momentum transfer range $0 \le Q^{2}\lesssim$  $4-5$  $GeV^{2}$ (up to $\sim 12$ $ GeV^{2}$ for the JLab upgrade). A major challenge for theory is that a quantitative description of the transition amplitudes must include soft nonperturbative contributions as well. Holographic models are good in that sence they have no limitation on $Q^{2}$. 

The ground state, $J^{P}=(1/2)^{+}$ with radial quantum number is  $n=0$ and angular momentum $l=0$; first excited state, $J^{P}=(1/2)^{-}$ with $(n,l)=(0,1)$; second excited state, $J^{P}=(1/2)^{+}$ $(n,l)=(1,0)$.

Recently the electromagnetic nucleon-Roper transition has been studied in the framework of soft-wall AdS/QCD model [12-14]. $N+\gamma^{\ast}\longrightarrow R(1440)$ transition was studied in Refs. [15-16] within the soft-wall model at the finite and zero temperature  cases. Electromagnetic transition form factors of nucleons were studied in the light-front holographic QCD in Refs. [17-19]. In Ref. [15] the Roper electroproduction was considered in a soft-wall AdS/QCD model [13,15,16,20,21] with inclusion of the leading three-quark (3q) state and higher Fock components, as well.
AdS/QCD models have other advantage is that they contain the correct power scaling description of form factors and helicity amplitudes at large $Q^{2}$  [22, 23]. These models also provide agreement with data at low and intermediate values of the momentum transfer $Q^{2}$ for these functions. The nucleon-Roper electromagnetic transition  within the light front holographic QCD was studied in Ref.  [12], where the Dirac form factor was considered. In Ref. [14] authors extended the soft-wall model of AdS/QCD to the description of all nucleon resonances with adjustable quantum numbers and considered form factors, helicity amplitudes and charge radii for the transition. In Ref. [24] authors showed that the in the framework of the extended version of the effective action the electromagnetic form factors of the nucleon and of the Roper resonance are sufficiently improved. This was achieved by including non-minimal interaction terms into the Lagrangian consistent with gauge invariance, which contributes to the momentum dependence of the form factors and helicity amplitudes. Moreover, in Ref. [14] it was presented a description of electromagnetic properties of the nucleon and the Roper at small finite temperatures using the formalism developed in Ref. [25]. In the [25] the authors present the study of the nucleon resonance state $N^{*} (1535)$ which has a negative-parity. In Ref. [19] the $N^{*}(1520)$ and $ N^{*}(1535)$  wave functions were defined without any adjustable parameters and are used to make predictions for the valence quark contributions to the transition form factors in semirelativistic aproximation. Tthe results are compared to data particularly for high momentum transfer. Authors study the structure of the $N^{*}(1710)$  resonance and calculate the $N+\gamma^{\ast}\longrightarrow N^{*}(1710)$ electromagnetic form factors, helicity amplitudes, and predict that they are almost identical to those of the $N+\gamma^{\ast}\longrightarrow R(1440)$  transition in the high momentum transfer  in covariant spectator quark model in Ref. [26]. In Ref. [27]  has been presented a comprehensive review of the electromagnetic transition form factors of baryons, where  the nucleon excitations $ N^{\ast}$ ( $\gamma^{\ast}N\longrightarrow N^{\ast}$) are in the first, second and third nucleon resonance states.  Despite some good results for the abovementioned transitions within the soft-wall model, to check holographic models for a form factors calculation, it is reasonable to consider transition form factors in the hard-wall model framework as well because, in several other form factor cases, the results of these models agree one with another [2, 28] and, in some cases, they differ [29].

	Here, we aim to consider the $N+\gamma^{\ast}\longrightarrow N^{*}(1440,1710)$  transitions with positive parity excited states and $N+\gamma^{\ast}\longrightarrow N^{*}(1535)$ transition with negative parity excited state and study the Dirac-Pauli electromagnetic and electric, magnetic Sachs  form factors, helicity amplitudes in the framework of hard-wall model.

	This paper is structured as follows: In Secs.  II and III we introduce bulk-to-boundary propagator for the electromagnetic field and the profile function for the nucleons. The electromagnetic current of the excited nucleon transition is defined in Sec. IV and in Sec. V, we discuss the form factors, charge and magnetic radii and helicity amplitudes of the $
	N+\gamma^{\ast}\longrightarrow N^{*} $ transition. In Sec. VI, we present our numerical analysis and results and comment these results in VII section.

\section{Vector field}
The flavour symmetry group the model is $SU(2)_{L} \times SU(2)_{R}$ subgroups of this group contain $A_{L}$ and $A_{R}$ gauge fields, correspondingly. Action for the model, which includes the pseudoscalar field $X$ is given:
\begin{equation}
S=\int  d^5x\sqrt{g}Tr\left \{ \left | DX \right |^{2}+3\left | X \right |^{2}-\frac{1}{4g_{5}^{2}}\left ( F_{L}^{2}+F_{R}^{2} \right )\right \}.
\label{1}
\end{equation}
Here $D_{\mu}X=\partial_{\mu}X-iA_{L\mu}X+iXA_{R\mu}$, $A_{L,R}=A_{L,R}^{a}t^{a}$, and $F_{\mu \nu}=\partial_{\mu}A_{\nu}-\partial_{\nu}A_{\mu}-i[A_{\mu},A_{\nu}]$. The constant $g_{5}$ was found in Ref. [30] by comparison of  two-point correlation function expressions for vector field obtained in QCD sum rules and hard-wall holographic model: $g_{5}^{2}=\frac{12\pi^{2}}{N_{c}}$. 
The expectation value of the field $X$ is determined by
the classical solution satisfying the UV boundary condition $(2/\epsilon)X(\epsilon) = M$ for quark mass matrix $M$ [30]:
\begin{equation}
X_{0}(z)=\frac{1}{2}Mz+\frac{1}{2}\Sigma  z^{3}.
\label{2}
\end{equation}
However, following to Refs. [13-15, 19] we ignore the chiral symmetry breaking term in the equation of motion for the spinor field. 
Introducing the vector field as $V = (A_{L} + A_{R})/2$ in the bulk, action for the vector field $V_{M}$ is written in the form:
		\begin{equation}
		S_{V}={-\frac{1}{2g_{5}^{2}}\int d^5x\sqrt{g}Tr{V_{MN}^{2}}},
		\label{3}
	\end{equation}
	where  $V_{MN}=\partial_{M}V_{N}-\partial_{N}V_{M}$ has been defined up to motion 	quadratic order in the action.  In the $V_{z}(x, z) = 0$ gauge this vector field corresponds to the photon field in the boundary theory. Fourier transform of the vector field $V_{\mu}$ is written as in the form Ref. [24]:
	\begin{equation}
	V_{\mu}(x,z)=\int\frac{d^4 q}{(2\pi)^{4}}e^{-iqx}V_{\mu}(q) V(q,z).
	\label{4}
	\end{equation}
Here  $V_{\mu}(q)$  is the UV boundary value of the  vector field. The equation of motion for the transverse part of the vector field  is obtained from the action  (\ref{3}) and was found in [30]:	
	\begin{equation}
		\partial_{z}\left (\frac{1}{z}\partial_{z}V(q,z)\right )+\frac{q^{2}}{z}V(q,z)=0.
		\label{5}
	\end{equation}
This equation was imposed the boundary conditions $V(q,\epsilon)=1$ and $\partial_{z}V(q,z_{m})=0$. As we consider the virtual photon in the boundary, we take $q^{2} \neq 0$ 	eigenvalues of the vector field.
 
 The solution to the equation (\ref{5}) is written in terms of the Bessel function [31,32]:
	\begin{equation}
		V(q,z)=\frac{\pi}{2}zq \left (\frac{Y_{0}(qz_{m})}{J_{0}(q,z_{m})}J_{1}(qz)+Y_{1}(qz) \right ).
		\label{6}
	\end{equation}
	Note that the vector field function $V_{\mu}(q)$ is the source for the nucleon current
	$J_{\mu}^{a}=\overline{N}\gamma_{\mu}t^{a}N$.
In the limit $Q^{2}\longrightarrow0$ the bulk-to-boundary propagator of the vector field can be written as in Ref. [33]
\begin{equation}
V(q,z)=1-\frac{Q^{2}z^{2}}{4}\left(1-2ln\left( \frac{z}{z_{m}} \right)  \right), \label{7}
\end{equation}
and in this limit the derivative $\partial_{z}V(q,z)$ accepts a form below:
\begin{equation}
\partial_{z}V(q,z)=Q^{2}zln\left( \frac{z}{z_{m}} \right). \label{8}
\end{equation}

\section{ Nucleons in hard-wall model}
	Electromagnetic transition form factors are an important probe for the study of the internal structure of nucleons. These form factors were studied in the framework of different models [16, 18, 26, 34]. We apply the 	hard-wall model of the holographic QCD for the calculation of $N^{*}\longrightarrow N$ transition form factors. The metric for this model is the five-dimensional AdS metric in Poincare patch [30]:
	\begin{equation}
		ds^{2}=\frac{1}{z^{2}}(\eta_{\mu\nu}dx^{\mu}dx^{\nu}-dz^{2}).
		\label{9}
	\end{equation}
	The fifth coordinate $z$ extends from $ 0 $ to $z_{m}$. These boundaries of the AdS space are called the ultraviolet (UV) and infrared (IR) ones, respectively.
	$ \eta_{\mu\nu} $ is the metric of the Minkowski space $ (\eta_{\mu\nu}=diag(1,-1,-1,-1);  \mu,\nu=0,1,2,3).$
	
	In order to describe nucleons in the AdS/CFT correspondence framework, it is necessary  to introduce two independent fermion fields in the bulk of AdS space, which describe the chiral components of the nucleons as is known from Ref. [35] (A further elaboration of this model can be found in Ref. [36]).  The minimal bulk action for the spinor field is written in the following:
	\begin{equation}
		S=\int dz
		d^4x\sqrt{g}\left [   i\bar{N_{1}}e^{M}_{A}\Gamma^{A}D_{M}N_{1}-m_{5}\bar{N_{1}}N_{1} + (1 \leftrightarrow 2 \& m_{5} \leftrightarrow - m_{5}) \right ]           ,
		\label{10}
	\end{equation}
	where $g$-is the determinant of the AdS metric, $ e^{M}_{A }$ is
	the vielbein for the metric (\ref{1}) and
	$\Gamma^{A}=(\gamma^{\mu},-i\gamma^{5}) $ are the Dirac matrices. $m_{5}=\pm\frac{5}{2}$ for $N_{1,2}$ correspondingly [35]. The Lorentz and gauge-covariant derivative is defined:
	\begin{equation}
		D_{M}=\partial_{M}-\frac{i}{4}\omega_{M}^{AB}\Sigma_{AB},
		\label{11}
	\end{equation}
	where $\omega_{M}^{AB}$ is the spin connection and $\Sigma_{AB}$ is defined as:  $\Sigma_{AB}=\frac{1}{2!}[\Gamma^{A},\Gamma^{B}]$.
	Nonvanishing components of the spin connections for the metric (\ref{9}) are the  following ones: $\omega_{\mu}^{5A}=-\omega_{\mu}^{A5}=\frac{1}{z}\delta_{\mu}^{A}$.
	Equation of motion obtained from the action (\ref{10})  has an explicit form:
	\begin{equation}
		(z\gamma^{5}\partial_{z}+iz\not{\!}{\partial}-2\gamma^{5})N_{1}-m_{5}N_{1}=0.
		\label{12}
	\end{equation}
	It is favorable to solve this equation in terms of Fourier components:
	\begin{equation}
		N_{1,2}=\frac{1}{2\pi}\int[\phi_{1,2L}(p,z)\psi_{1,2L}(p)+\phi_{1,2R}(p,z)\psi_{1,2R}(p)]e^{-ipx}d^4p, \label{13}
	\end{equation}
	where the $4D$ spinors satisfy the free Dirac equation:
	\begin{equation}
		\not{\!}{p}\Psi(p)=|p|\Psi(p). \label{14}
	\end{equation}
	Here $p$ denotes $p=|p|=\sqrt{p^{2}}$  for a time-like
	four-momentum $p$. Then the Dirac equation (\ref{12}) will be reduced to equations for the profile functions $\phi_{1,2L,R}$ [35]:
	\begin{eqnarray}
		\left(\partial_{z}^{2}-\frac{4}{z}\partial_{z}+\frac{6+m_{5}-m_{5}^{2}}{z^{2}}\right)\phi_{1,2L}\left(pz\right)=-p^{2}\phi_{1,2L}\left(pz\right),\nonumber \\
		\left(\partial_{z}^{2}-\frac{4}{z}\partial_{z}+\frac{6-m_{5}-m_{5}^{2}}{z^{2}}\right)\phi_{1,2R}\left(pz\right)=-p^{2}\phi_{1,2R}\left(pz\right). \label{15}
	\end{eqnarray}
	Extra components of these spinors are eliminated by the  UV and IR boundary conditions [3] and normalizability analysis. The mass eigenvalues corresponding to the  Kaluza-Klein modes are determined by the IR boundary condition $\phi_{R}(z_{m})=0$. Solutions to the equations (\ref{15}) are expressed in terms of Bessel functions $J_{2,3}$ [35, 37]:
	\begin{eqnarray}
		\phi_{1L}^{n}\left(pz\right)=-c_{1}^{n}z^{\frac{5}{2}}J_{2}\left(pz\right),
		\phi_{1R}^{n}\left(pz\right)=c_{1}^{n}z^{\frac{5}{2}}J_{3}\left(pz\right),\nonumber \\
		\phi_{2L}^{n}\left(pz\right)=-c_{2}^{n}z^{\frac{5}{2}}J_{3}\left(pz\right),
		\phi_{2R}^{n}\left(pz\right)=c_{2}^{n}z^{\frac{5}{2}}J_{2}\left(pz\right),
		\label{16}
	\end{eqnarray}
	where $c_{1,2}^{n}$ are constants were found from the normalization  conditions and are equal [37]:
	\begin{equation}
		|c_{1,2}^{n}|=\frac{\sqrt{2}}{z_{m}J_{2}(M_{n}z_{m})}.
		\label{17}
	\end{equation}
	Here, $ M_{n}$ is the mass spectrum of the Kaluza-Klein modes
	corresponding to the ground and excited states of the boundary
	nucleons. The spectrum $M_{n}$ is expressed in terms of zeros
	$\alpha_{n}^{(3)}$ of the Bessel function $J_{3}$:
	\begin{equation}
		M_{n}=\frac{\alpha_{n}^{(3)}}{z_{m}}. \label{18}
	\end{equation}

	\section{The electromagnetic  current}
	Electromagnetic transition form factors for the nucleon to excited nucleon
	transition have been studied in several theoretical approaches. These 
	form factors  are defined due to Lorentz and gauge invariance of the
	interaction by the following matrix element [24, 38]:
	\begin{equation}
		M^{\mu}(p_{1},\lambda_{1},p_{2},\lambda_{2})=\overline{u}(p_{1},\lambda_{1})
		[\gamma_{\bot}^{\mu}F_{1}(Q^{2})+i\sigma^{\mu\nu}\frac{q_{\nu}}{M}F_{2}(Q^{2})]u(p_{2},\lambda_{2}),
		\label{19}
	\end{equation}
	where $\overline{u}\left(p_{1},\lambda_{1}\right)$ and
	$u\left(p_{2},\lambda_{2}\right)$  are the spinors describing the
	nucleon resonance and the nucleon, respectively. $M$ is the sum of
	the nucleon and excited nucleon masses $ M= M_{N}+M_{N^{*}}$; $\gamma_{\bot}^{\mu}=\gamma^{\mu}-q^{\mu}\frac{q}{q^{2}}$,
	$q=p_{1}-p_{2}$. The final and initial nucleons helicities
	$(\lambda,\lambda_{1})$ are related to the
	$\lambda_{2}=\lambda_{1}-\lambda$ photon helicity. The
	four-momenta of $N^{*}$ nucleon, nucleon and photon and the polarization vector
	of a photon are specified as [39, 40]:
	\begin{equation}
		p_{1}^{\mu}=(M_{1}, \textbf{0}),   p_{2}^{\mu}=(E,0,0,-|\textbf{p}|),
		q^{\mu}=(q^{0},0,0,|\textbf{p}|)] \label{20}
	\end{equation}
	\begin{equation}
		\epsilon^{\mu}(\pm)=(0,\vec{\epsilon}^{\pm}) ,
		\vec{\epsilon}^{\pm}=\frac{1}{\sqrt{2}}(\pm1,i,0),
		\label{21}
	\end{equation}
	\begin{equation}
		\epsilon^{\mu}(0)=\frac{1}{\sqrt{Q^{2}}}(|\textbf{p}|,0,0,q^{0}),
		\label{22}
	\end{equation}
	where  $E=\frac{Q_{+}}{2M_{N^{*}}}-M_{N}$,
	$Q_{\pm}=M_{\pm}^{2}+Q^{2}$, $Q^{2}=-q^{2}$, $M_{\pm}=M_{N^{*}}\pm
	M_{N}$ and $|\textbf{p}|=\frac{\sqrt{Q_{+}Q_{-}}}{2M_{N^{*}}}$ is the value of
	the three-momentum of the nucleon or the photon.  For the positive parity resonance state the electromagnetic current of the  excited nucleon-nucleon transitions is defined as below [41-44]:
	\begin{equation}
		J^{\mu}=\overline{u}_{f}(p_{f})[\gamma_{\mu}^{T}F_{1}^{fi}(Q^{2})+\frac{1}{m_{fi}}\sigma_{\mu\nu}Q_{\nu}F_{2}^{fi}(Q^{2})]u_{i}(p_{i}).
		\label{23}
	\end{equation}

	Here, $p_{i,f}$ are four-momenta of the  incoming/outgoing nucleons: $p_{i,f}^{2}=m_{i,f}^{2}$; $Q=p_{f}-p_{i}$; $m_{fi}=m_{f}+m_{i}$. $F_{1,2}$ are called Dirac and Pauli form factors, respectively.

	\section{Dirac and Pauli form factors and helicity amplitudes in AdS/QCD}
	\subsection{Form factors}
	The  action for the bulk interaction between the fields  is defined as:
	\begin{equation}
		S_{int}=\int d^4 x dz\sqrt{g}L_{int}(x,z), \label{24}
	\end{equation}
	where $L_{int}(x,z)$ is the Lagrangian of the interaction between the spinor and vector fields [13,24]:
	\begin{equation}
		L_{int}(x,z)= c_{\tau}^{N^{*}N}\overline{\Psi}^{N^{*}}_{i}(x,z)\hat{V_{i}}(x,z)\Psi_{i}^{N}(x,z).
		\label{25}
	\end{equation}
	Here $c_{\tau}$ is the mixing parameters that contribute of the AdS fermion fields with twist dimension $3$ [24].  $\overline{V}_{\pm}(x,z)$ denotes :
	\begin{equation}
		\overline{V}_{\pm}(x,z)=\tau_{3}\Gamma^{M}V_{M}(x,z)\pm\frac{i}{4}\eta_{V}[\Gamma^{M},\Gamma^{N}]V_{MN}(x,z)\pm g_{V}\tau_{3}\Gamma^{M}i\Gamma^{z}V_{M}(x,z).
		\label{26}
	\end{equation}
	
As seen from the (\ref{24})  formula the coupling and electromagnetic field $V_{M}$  include to the action of the  electromagnetic interaction with the hadrons. $\Gamma^{M}V_{M}(x,z)$ is the minimal coupling where  $\tau_{3}$ is the isospin Pauli matrix and $\Gamma^{M}$ is a $5D$ gamma matrix. With the minimal
Dirac coupling one obtains only contributions for
the Dirac form factor [18]. $\frac{i}{4}\eta_{V}\left [\Gamma^{M},\Gamma^{N}\right ] V_{MN}(x,z)$ term is included to the non-minimal coupling for generating contributions of the Pauli form factor where $\eta_{V}$ is the matrix of the $\eta_{p},\eta_{n}$  coupling constants: $\eta_{V}=diag(\eta_{p},\eta_{n})$ and is associated nucleon anomalous magnetic moment, and $V_{MN}=\partial_{M}V_{N}-\partial_{N}N_{M}$.  This non-minimal coupling gives also an extra contribution for the Dirac form factor [33]. $g_{V}\tau_{3}\Gamma^{M}i\Gamma^{z}V_{M}(x,z)$ is the minimal-type coupling where $g_{V}$ is an isovector coupling constant  [13, 19]. In Ref. [21] the authors showed that $ c_{\tau}^{N^{*}N}$ are constrained by the condition $\sum_{\tau}c_{\tau}^{N^{*}N}=1$ in order to get the correct normalization of the kinetic term of the four-dimensional spinor field, and is consistent with
electromagnetic gauge invariance, as well as.

	 After taking into account  the definitions  (\ref{25}), (\ref{26}) in (\ref{24}), $S_{int}$ gets following explicit form:
	\begin{eqnarray}
		S_{int}=\int\int\frac{dz}{z^{5}}d^4 p d^4
		p^{\prime}V(q,z)V_{\mu}(q)\left[\frac{c_{\tau}^{N^{*}N}z}{2}\left[\phi_{L}^{\ast}(z)\phi_{L}(z)+\phi_{R}^{\ast}(z)\phi_{R}(z)+\phi_{R}^{\ast}(z)\phi_{R}(z) \right.\right. \nonumber \\
		\left.\left.-\phi_{L}^{\ast}(z)\phi_{L}(z)\right]\overline{u}(p^{\prime})\gamma^{\mu}u(p)-\eta_{V}z^{2}q_{\nu}\left[\phi_{R}^{\ast}(z)\phi_{R}(z)
		-\phi_{L}^{\ast}(z)\phi_{L}(z)  \right. \right. \nonumber \\
		\left.\left.
		-\phi_{R}^{\ast}(z)\phi_{L}(z)-\phi_{L}^{\ast}(z)\phi_{R}(z)\right]\overline{u}(p^{\prime})\sigma^{\mu\nu}u(p)\right].
		\label{27}
	\end{eqnarray}
	Here $|p|=M_{nuc}$, $|p^{\prime}|=M_{exc.}$ and $q_{\nu}=p-p^{'}$. The profile
	functions $\phi_{L,R}^{\ast}\left(|p^{\prime}|,z\right)$ and $\phi_{L,R}\left(|p|,z\right)$ describe the excited state and nucleon, respectively.
	According to AdS/CFT correspondence of the bulk and boudary theories the generating functionals of the gauge and AdS gravity theories are equivalent:
	\begin{equation}
		Z_{gauge}=Z_{AdS}. \label{28}
	\end{equation}
	The vector current of nucleons, which is defined in the QCD theory, can be calculated from the $Z_{AdS}$ using the correspondence equality (\ref{28}). The vacuum expectation value of the nucleon's vector current $J_{\mu}$ in the boundary QCD theory  will be found applying the holographic formula:
	\begin{equation}
		\langle J_{\mu}\rangle=-i\frac{\delta Z_{QCD}}{\delta V_{\mu}(q)}|_{V_{\mu}=0}=-i\frac{\delta e^{iS_{int}}}{\delta V_{\mu}(q)}|_{V_{\mu}=0}.    \label{29}
		\end{equation}
		
			From the comparison the electromagnetic currents of the excited nucleon-nucleon  transitions (\ref{23}) with the nucleon vector current (\ref{29}) for the $Q^{2}=-q^{2}>0$ momentum transfer region  the holographic expressions of the $G_{i}(Q^{2})$ form factors  will be written in terms of integrals over the $z$ variable:
			\begin{equation}
				G_{1}(Q^{2})=\frac{1}{2}\int_{0}^{z_{m}}dzV(Q,z)c_{\tau}^{N^{*}N}(\phi_{L}(m_{N},z)\phi_{L}(m_{N^{*}},z)+\phi_{R} ({m_{N}},z)\phi_{R}(m_{N^{*}},z)), \label{30}
			\end{equation}
			\begin{equation}
				G_{2}(Q^{2})=\frac{1}{2}\int_{0}^{z_{m}}dzV(Q,z)c_{\tau}^{N^{*}N}(\phi_{R}(m_{N},z)\phi_{R}(m_{N^{*}},z)-\phi_{L}(m_{N},z)\phi_{L}(m_{N^{*}},z)), \label{31}
			\end{equation}
			\begin{equation}
				G_{3}(Q^{2})=\frac{1}{2}\int_{0}^{z_{m}}dz\partial_{z}V(Q,z)c_{\tau}^{N^{*}N}(\phi_{L}(m_{N},z)\phi_{L}(m_{N^{*}},z)-\phi_{R}(m_{N},z)\phi_{R}(m_{N^{*}},z)),  \label{32}
			\end{equation}
			\begin{equation}
				G_{4}(Q^{2})=\frac{M}{2}\int_{0}^{z_{m}}dzV(Q,z)c_{\tau}^{N^{*}N}(\phi_{L}(m_{N},z)\phi_{R}(m_{N^{*}},z)+\phi_{L}(m_{N^{*}},z)\phi_{R}(m_{N},z)).  \label{33}
			\end{equation}
			The Dirac and Pauli form factors $F_{1}(Q^{2})$ and $F_{2}(Q^{2})$  are defined by means  of the $G_{i}(Q^{2})$ form factors [16]:
			\begin{equation}
				F_{1}(Q^{2})=G_{1}(Q^{2})+g_{V}G_{2}(Q^{2})+\eta_{V}G_{3}(Q^{2}),  \label{34}
			\end{equation}
			\begin{equation}
				F_{2}(Q^{2})=\eta_{V}G_{2}(Q^{2}) .  \label{35}
			\end{equation}
			
			The $F_{1,2}(Q^{2})$ form factors are normalized to the electric charge $e_{N}$ and anomalous magnetic moment $\mu_{a}$ of the  proton: $F_{1}(0)=e_{N}$ and $F_{2}(0)=\mu_{a}=g(e/2M)=1.79 \mu_{B}$. A point particle of charge $e$ and total magnetic moment $ (g+1)\mu_{B}$ is a particle for which $F_{1}(Q^2)=e$ and $F_{2}(Q^2)=g\mu_B$ for all values of $ Q^2$. The functions $F_{1}(Q^{2})$ and $F_{2}(Q^{2})$ are called the charge and magnetic moment form factors of the nucleon, respectively.

The electric and magnetic Sachs form factors of the nucleon $G_{E}(Q^{2})$, $G_{M}(Q^{2})$ are alternative Lorentz invariant quantities, and are defined in terms of  Dirac and Pauli form factors within the relations: 
\begin{equation}
G_{E}(Q^{2})=F_{1}(Q^{2})-\frac{Q^{2}}{4m_{N}^{2}}F_{2}(Q^{2}),  \label{36}
\end{equation}
\begin{equation}
G_{M}(Q^{2})=F_{1}(Q^{2})+F_{2}(Q^{2})  .   \label{37}
\end{equation}
The slopes of the form factors in the limit $Q^{2}\longrightarrow 0$ are defined as the electric and magnetic charge radii of the nucleon:	
\begin{equation}
\left\langle r_{E}^{2} \right\rangle =-6\frac{dG_{E}(Q^{2})}{dQ^{2}}|_{Q^{2}=0} ,    \label{38}
\end{equation}	
\begin{equation}
\left\langle r_{M}^{2} \right\rangle ={-6}{G_{M}(0)}\frac{dG_{M}(Q^{2})}{dQ^{2}}|_{Q^{2}=0} .  \label{39}
\end{equation}
			\subsection{Helicity amplitudes}
			In addition to the electromagnetic transition form factors the  helicity amplitudes $A_{1/2}(Q^{2})$ and $S_{1/2}(Q^{2})$ are defined through the transition matrix element of the transverse electromagnetic interaction between the nucleon and the nucleon resonance states:
			\begin{eqnarray}
				A_{1/2}(Q^2)=\sqrt{\frac{2\pi\alpha}{K}}\left\langle R,S_{z}^{'} =+\frac{1}{2} \left| \varepsilon_{+} J  \right| N,  S_{z}=-\frac{1}{2} \right\rangle,  \label{40}
			\end{eqnarray}
			\begin{eqnarray}
				S_{1/2}(Q^2)=\sqrt{\frac{2\pi\alpha}{K}}\left\langle R,S_{z}^{'} =+\frac{1}{2} \left| \varepsilon_{0} J  \right| N, S_{z}=-\frac{1}{2} \right\rangle \frac{\left|\textbf{q} \right|}{Q},\label{41}
			\end{eqnarray}
			where $\textbf{q}$ is the photon three-momentum in the rest  frame of nucleons, $\varepsilon_{\lambda}^\mu(\lambda=0,\pm1)$ is the photon polarization vector, $\alpha \simeq 1/137$ is the fine-structure constant, $K=\left|\textbf{q} \right|=\frac{M_{N^{*}}^2-M_{N}^2}{2M_{N^{*}}}$. The  magnitude of  the  photon  three-momentum is defined:
			\begin{equation}
			 \left| \textbf{q} \right|=\frac{\sqrt{Q_{+}^2Q_{-}^2}}{2M_{N^{*}}},  \label{42}
			\end{equation}
			where $Q_{\pm}=(M_{N^{*}}\pm M_{N})^{2}+Q^{2} $. Note that when $Q^{2}=0$, one has $K=\left| \textbf{q} \right|=\frac{M_{N^{*}}^{2}-M_{N}^{2}}{2M_{N^{*}}}$.   
			For the  $ N+\gamma^{\ast}\longrightarrow N^{*}(1440,1710)$ transition, where excited state is the $J^{P}=+1/2$ state, the $A_{1/2} (Q^2)$, $S_{1/2}(Q^2)$  helicity amplitudes are written in terms of the  Dirac and  Pauli  form factors as following [34, 45]: 
			\begin{equation}
				A_{1/2}(Q^2)=R[F_{1}(Q^2)+\frac{M_{N^{*}}-M_{N}}{M_{N^{*}}+M_{N}}]F_{2}(Q^2),   \label{43}
			\end{equation}
			\begin{equation}
				S_{1/2}(Q^2)=\frac{R}{\sqrt{2}} \left| \textbf{q} \right|\frac{M_{N^{*}}+M}{Q^2}(F_{1}(Q^2)-\tau F_{2}(Q^2)),  \label{44}
			\end{equation} 		
	where $\tau$ and $R$ denote: $\tau=\frac{Q^2}{(M_{N^{*}}+M)^2} $, $R=\sqrt{\frac{\pi \alpha Q_{-}^2}{M_{N^{*}}MK}}.$
 For the  $ N+\gamma^{\ast}\longrightarrow  N^{*}(1535)$ transition, where resonance state is the $J^{P}=-1/2$ state  the $A_{1/2} (Q^2)$, $S_{1/2}(Q^2)$ helicity amplitudes can be defined in terms of the $F_{1,2}(Q^2)$ form factors as below [19]: 
\begin{equation}
				A_{1/2}(Q^2)=2A_{R}[F_{1}(Q^2)+\left[ \frac{M_{N^{*}}-M_{N}}{M_{N^{*}}+M_{N}}\right] F_{2}(Q^2),   \label{45}
			\end{equation}
			\begin{equation}
				S_{1/2}(Q^2)=-\sqrt{2}A_{R}(M_{N^{*}}+M_{N})\frac{ \left| \textbf{q} \right|}{Q^{2}} \left[ \frac{M_{N^{*}}-M_{N}}{M_{N^{*}}+M_{N}}(F_{1}(Q^2)-\tau F_{2}(Q^2)) \right] .  \label{46}
			\end{equation} 	
	
	\section{Numerical analysis}
	We perform the numerical calculations for of the $F_{1,2}(Q^{2})$, $G_{E,M}(Q^{2})$ form factors and $A_{1/2}(Q^2)$, $S_{1/2}(Q^2)$ helicity amplitudes according to the (\ref{34}), (\ref{35}), (\ref{36}), (\ref{37}) and (\ref{43}), (\ref{44}), (\ref{45}), (\ref{46}) formulas. The parameters and constants in these formulas are fixed as $c_{3}^{N^{*}N}=0.72$, $\eta_{p}=0.453$    [13], $z_{m}= (0.205$ GeV $)^{-1}$  [37].  The main question of parameter fixing is how to fix the parameter of the holographic coordinate - the hard-wall cutoff, which is the only parameter of the model. In the model containing only one field, usually this parameter is fixed by  corresponding the first KK mode’s mass to the particle’s ground state mass known from the experiment. In such a way in different works it were found the following values for this parameter, for instance, $z_{m} =(0.205$ GeV $)^{-1}$  in Ref. [37], $z_{m} = (0.324 $ GeV $)^{-1}$  in Ref. [2], $z_{1} = (0.1473$ GeV $ )^{-1}$ in Ref. [2]. When the model contains two or more interacting fields, there are several physical quantities (besides two or more spectra) related to the interaction between the fields, and these quantities are measurable in the experiment. The model predictions for all quantities are sensitive to the value of $z_{m}$. For instance, in Ref. [2]  $z_{m} = (0.286 $ GeV $ )^{-1}$ value was chosen, to reach the coincidence of the model’s result for form factor graph to the experimental one, while another value of $z_{m}$ fits a mass spectrum $(z_{1}= (0.1473 $ GeV $)^{-1})$. Here, we investigate four form factors and two helicity amplitudes for three states and try to fit them to the experimental values. Of course, it is impossible to get results agreeing with the experimental data simultaneously for all of them by fixing only one parameter. Numerical analysis of all values above shows that the parameter value $z _{m} =( 0.205 $ GeV $)^{-1}$  is suitable for all dependencies, as all quantities have a $Q^{2}$ range that agrees with experimental data. Thus, we choose $z_{m} =( 0.205$ GeV $)^{-1}$ value to fix parameter.

	In Fig. 1, we present our numerical results  for the $ N+\gamma^{*}\longrightarrow R(1440) $  electromagnetic transition in the $0 \leq Q^2 \leq 5$ $ GeV^{2} $ interval. We plot the Dirac and Pauli form factors and compare our results to the MAID [40], CLAS [46] experimental data and valence quark contributions model [34]. As is seen from the Fig.1a the Dirac form factor for this transition is close to the data from CLAS and MAID experiments in the $1 \leq Q^2 \leq 5$ $ GeV^{2} $ interval. Plot for the Pauli form factor  is presented in the Fig. 1b and agrees with experimental data starting from $Q^{2}=0.9$  $ GeV^{2}$ value of momentum transfer. However, at low values of $Q^{2}$ $(Q^{2}< 1)$ $GeV^{2}$ the hard-wall model results don't describe data for this transition in a good agreement.
	
	In Fig. 2 the helicity amplitude $S_{1/2}( Q^2)$ presented for this transition. Though helicity amplitudes have same shape of dependence the hard-wall model results describe the experimental data with less accuracy than the valence quark contribution result. 

	In Figs. 3-4 we present plots for the form factors and helicity amplitudes for the transition with the negative parity sate $ N^{*}(1535)$ .  For this transition the plot for $F_{1}(Q^2)$ form factor (Fig. 3 left) is close to experimental data in the region $2\leq Q^2 \leq 4$ $ GeV^{2} $ and again at the low energy limit hard-wall model fails in description of transition. For the $F_{2}(Q^2)$ form factor (Fig. 3 right) the hard-wall results is better than semirelativistic approximation ones and close to experimental data in the $2\leq Q^2 \leq 4$ $ GeV^{2} $ region. The helicity amplitudes for the $ N^{*}(1535)$ transition $A_{1/2}(Q^2)$ and $S_{1/2}(Q^2)$ for the $ N^{*}(1535)$ transition are close to the data in the $Q^{2}\geq1$ $GeV^{2}$ region (Fig. 4).

	Graphs for the $ N^{*}(1710)$ transition are presented in Figs. 5 - 6. The form factors $F_{1}(Q^2)$ and $F_{2}(Q^2)$ agrees with data in the region $1 \leq Q^2 \leq 5$ $ GeV^{2} $. As seen from Fig. 6, neither the hard-wall model nor the non-relativistic quark one correctly describes the helicity amplitudes $A_{1/2}(Q^2)$ and $S_{1/2}(Q^2)$ in the $ N^{*}(1710)$ transition state case.

	Electric and magnetic form factors  $G_{E}(Q^{2})$ and $G_{M}(Q^{2})$ for the negative parity $ N^{*}(1535)$ state transition are given in Fig. 7. The hard-wall results for the   $G_{E}(Q^{2})$ form factor is close to MAID experimental data in the region $1 \leq Q^2 \leq 6$ $ GeV^{2} $. The $G_{M}(Q^{2})$ graph is close to CLAS experimental data in the interval  $1 \leq Q^2 \leq 6$ $ GeV^{2} $. Also, both graphs for $G_{E}(Q^{2})$, $G_{M}(Q^{2})$ form factors are close to semirelativistic approximation graphs in this interval. It is worth to notice that for this transition, the hard-wall model results for the $G_{E}(Q^2)$ and $G_{M}(Q^{2})$  are in a good  agreement with  experimental data. In Fig. 8 plots of the  $G_{E}(Q^{2})$ and $G_{M}(Q^{2})$ form factors were presented for the $R(1440)$ and $ N^{*}(1710)$ transitions which are in good agreement with experimental and light front holography data.
	
	In the last column of the table 1 were given radii for the unexcited nucleons. We have shown the differences between our results and experimental data for the unexcited nucleons are small and compared with the data. Also,  in this table comparison of soft-wall model results with the ground state nucleons show on small differences between them. This allows us to make conclusion that the radii have a little change on excitation.
	
\section{Summary}
We investigate nucleon electromagnetic form factors and helicity amplitudes within  the hard-wall model for the nucleon-excited nucleon transitions $ N+\gamma^{*}\longrightarrow  N^{*}(1440,1535,1710)$. We also check this model for the charge and magnetic radii of the nucleon in these transition cases. As is seen from the comparison of the graphs, the holographic hard-wall model gives good results for the Dirac and Pauli form factors and helicity amplitude behaviors in the $Q^{2}\geq1$ $GeV^{2}$ values of momentum transfer. Failure of the hard-wall model in the description of nucleon transition form factors at low momentum transfer values may be related to several reasons: 

1) Simplicity of the model in the sense that it doesn't take into account more complicated physical situation at low energies, i.e., it takes into account interaction of particles at low and high values of momentum transfer in the same way. But it may thought that transitions at high transfer momentum occur with higher probability than at low values of it;

2) It doesn't take into account interaction with the spins or magnetic moments of particles;

3) The study here doesn't include higher order derivative terms in the interaction Lagrangian. Also, in the hard-wall model containing two or more interacting fields, it is problematic to use the only parameter $z_{m}$ for fitting both the spectrum of the Kaluza-Klein modes to the particle's known mass spectrum and simultaneously to fit the value of studying physical quantity, with the experimental data for it [2]. 
 Despite this, the hard-wall model gives results describing the magnetic and charge form factors for the $ N^{*}(1440, 1710)$ excited state transition case well, and it is worth carrying out the phenomenological studies within the hard-wall model as well.

\newpage

\begin{figure}[!h]
    \begin{center}
        \includegraphics[width=6cm]{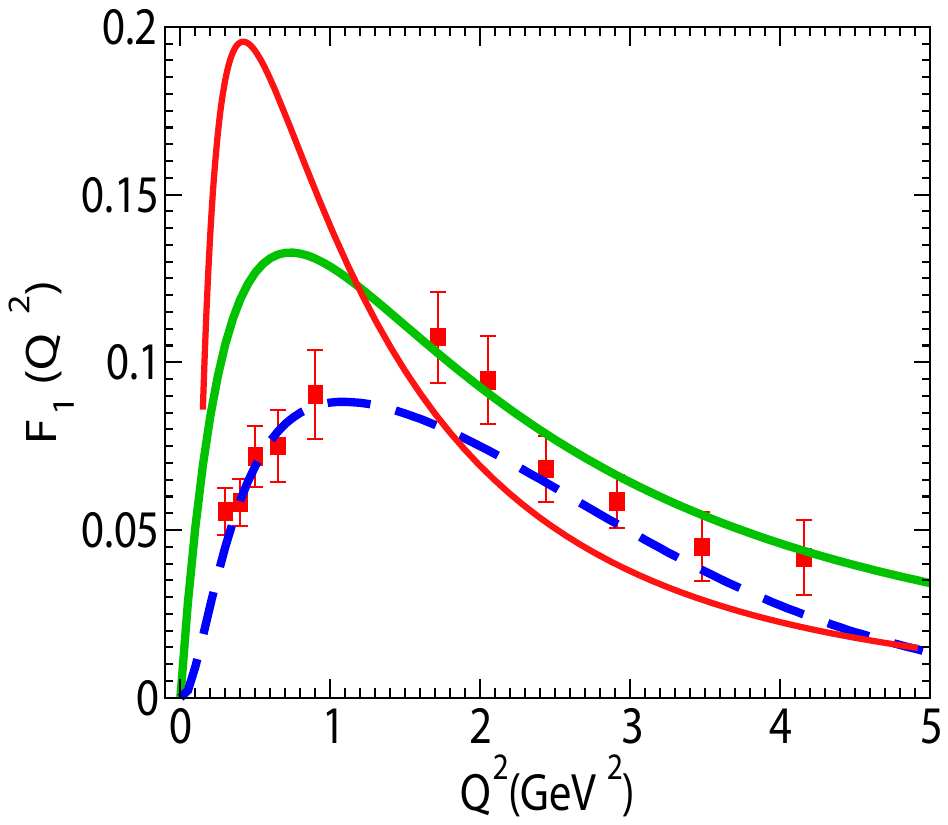}
        \includegraphics[width=6cm]{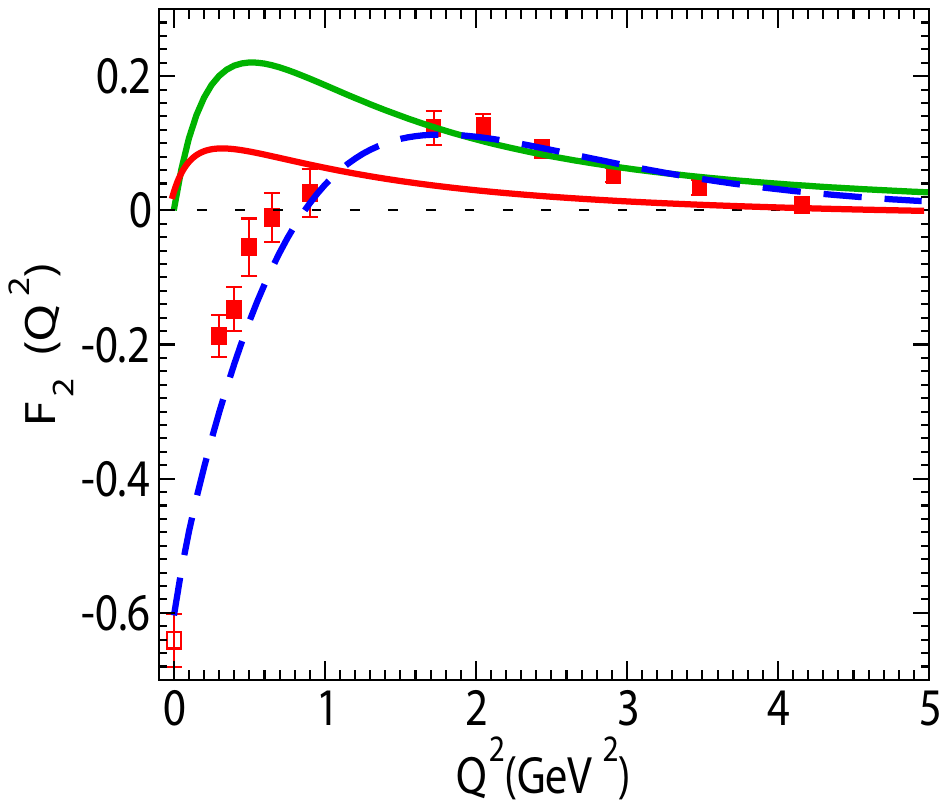}
    \end{center}
    \caption{ $N+\gamma^{\ast}\longrightarrow R(1440)$ transition Dirac and Pauli form factors (red lines) is compared with CLAS experimental data (squares with error bars) [46], MAID fit (dashed lines) [40] and valence quark contributions  (solid lines) [34].}
    \label{f2MB}
\end{figure}

\begin{figure}[!h]
    \begin{center}
        \includegraphics[width=6cm]{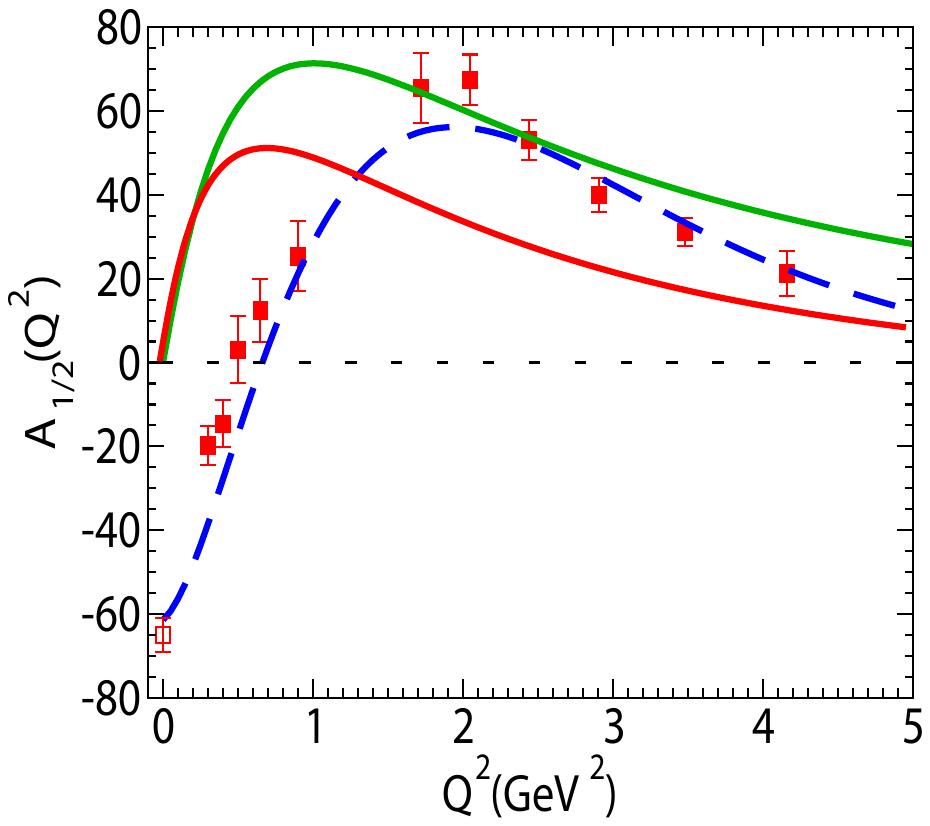}
        \includegraphics[width=6cm]{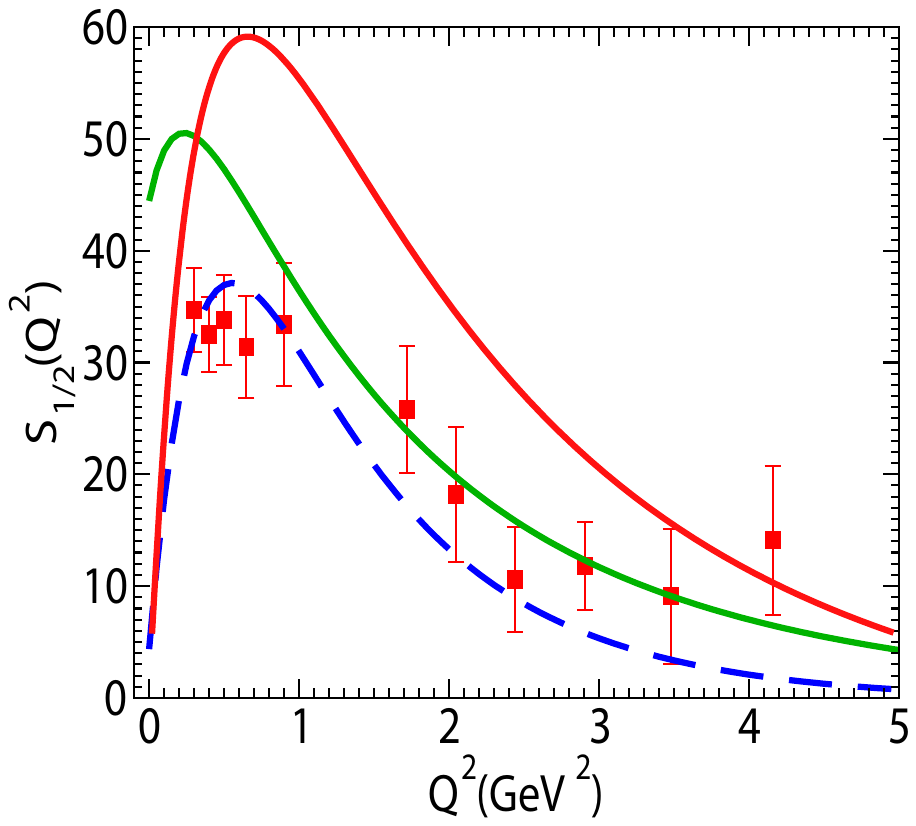}
    \end{center}
    \caption{ $N+\gamma^{\ast}\longrightarrow R(1440)$ transition $A_{1/2}(Q^2)$ and  $S_{1/2}(Q^2)$ helicity amplitudes in units of $ 10^{-3}$  $ GeV^{-1/2}$ (red lines) is compared with CLAS experimental data (squares with error bars) [46], MAID fit (dashed lines) [40]  and valence quark contributions  (green solid lines) [34] are also shown.}
    \label{f2MB}
\end{figure}

\begin{figure}[!h]
    \begin{center}
        \includegraphics[width=6cm]{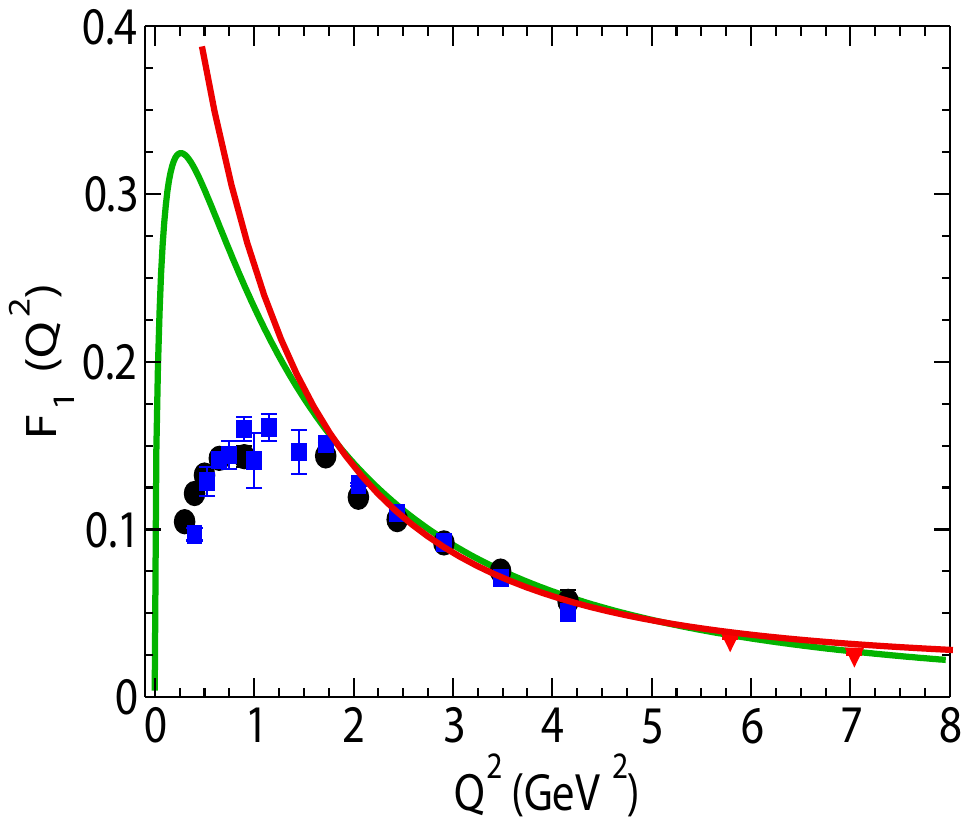}
        \includegraphics[width=6cm]{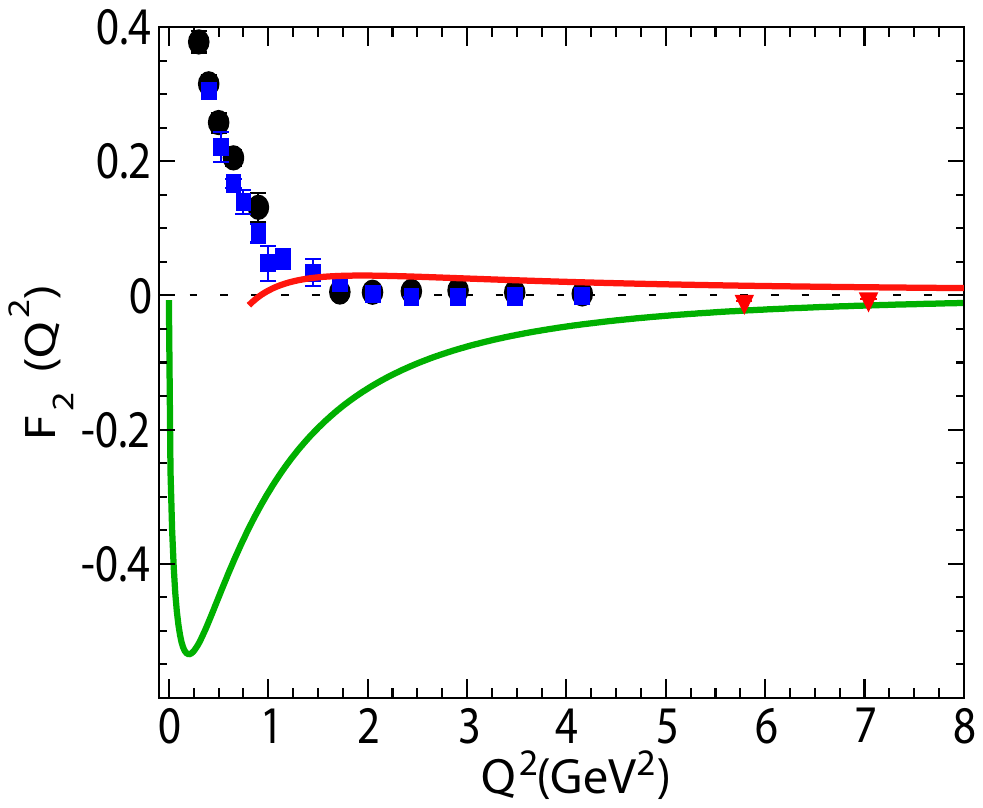}
    \end{center}
    \caption{ Results for the $N+\gamma^{*}\longrightarrow N^{*}(1535)$ transition Dirac and Pauli form factors (red lines) is compared with CLAS experimental data (full circles) [46], MAID (full squares) [47, 48], JLab/Hall C (triangles) [49] and the semirelativistic approximation  (green thick solid line) [19].}
    \label{f2MB}
\end{figure}

\begin{figure}[!h]
    \begin{center}
        \includegraphics[width=6cm]{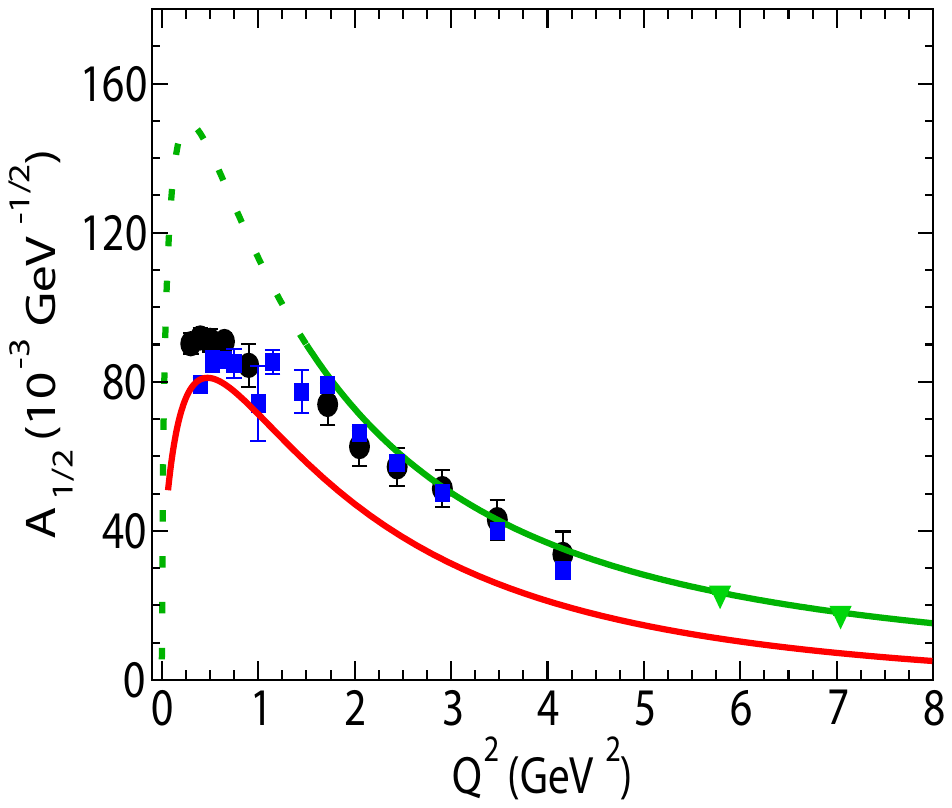}
        \includegraphics[width=6cm]{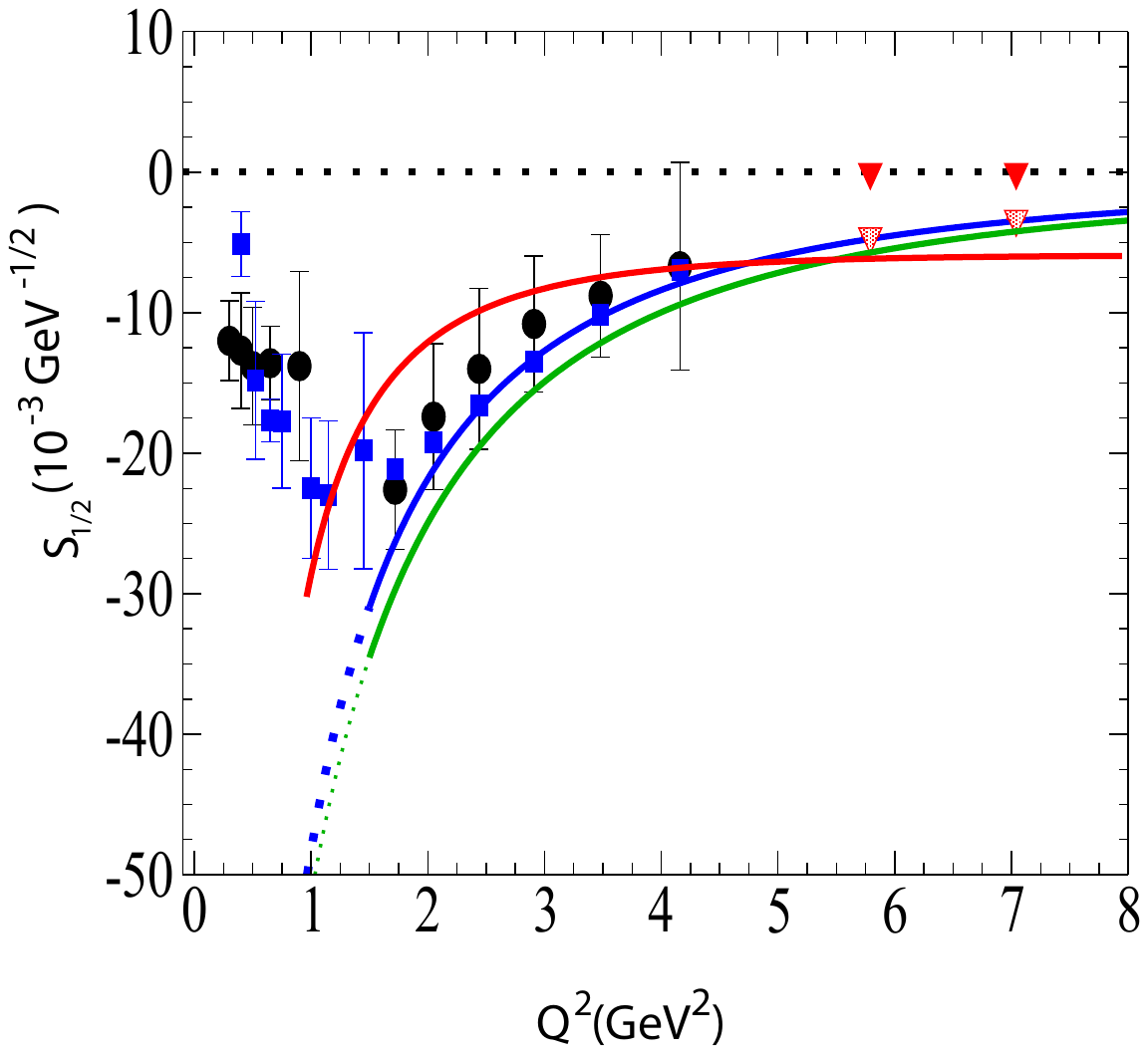}
    \end{center}
    \caption{Results for the $N+\gamma^{\ast}\longrightarrow N^{*}(1535)$ transition $A_{1/2}(Q^2)$ and  $S_{1/2}(Q^2)$ helicity amplitudes in units of $GeV^{-1/2}$ (red lines)  is compared with CLAS experimental data (full circles) [46], MAID (full squares) [47, 48], JLab/Hall C (triangles) [49], PDG (empty squares) [50] and the semirelativistic approximation  (green thick solid line) [19].}
    \label{f2MB}
\end{figure}

\begin{figure}[!h]
    \begin{center}
        \includegraphics[width=6cm]{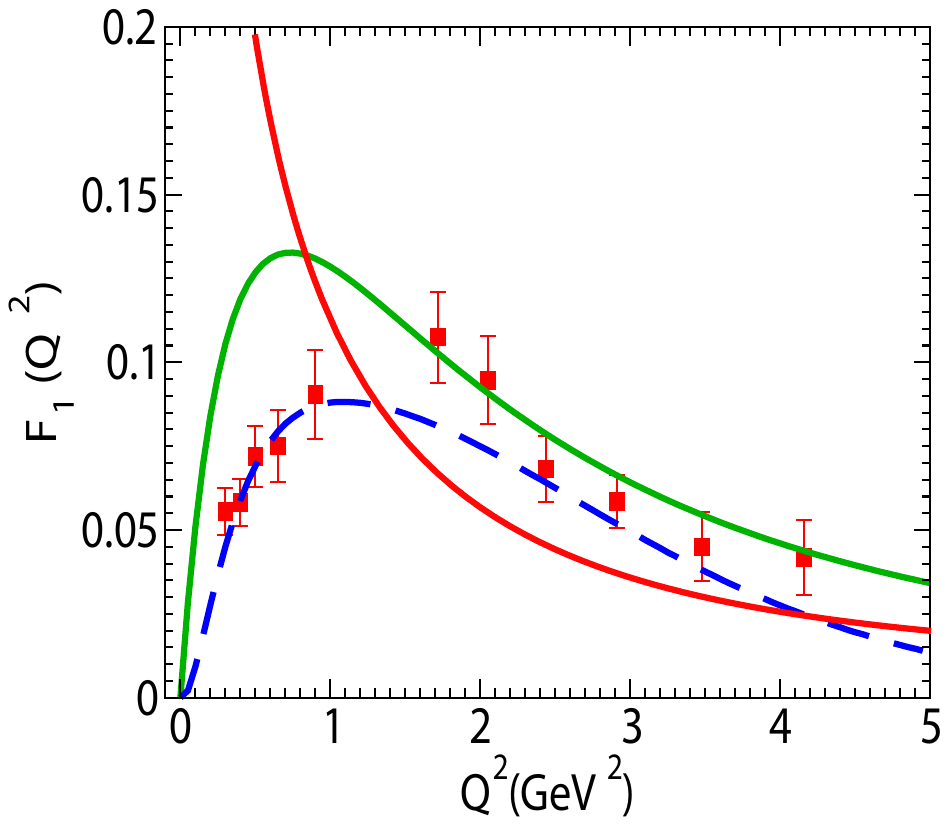}
        \includegraphics[width=6cm]{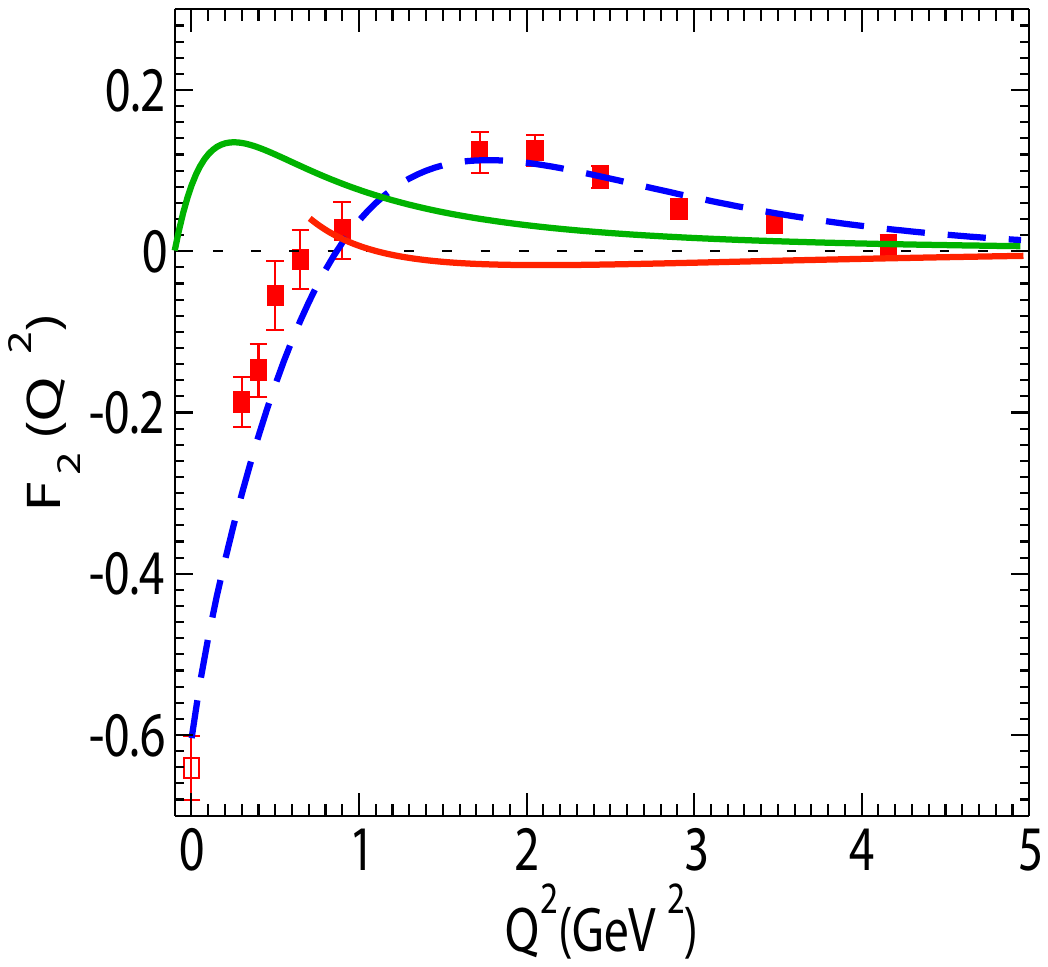}
    \end{center}
    \caption{$N+\gamma^{\ast}\longrightarrow N^{*}(1710)$ transition  Dirac and Pauli form factors (red lines) compared with CLAS experimental data(squares with error bars) [46], MAID fit (dashed lines) [40] and nonrelativistic quark model (green lines) [26].}
    \label{f2MB}
\end{figure}

\begin{figure}[!h]
    \begin{center}
        \includegraphics[width=6cm]{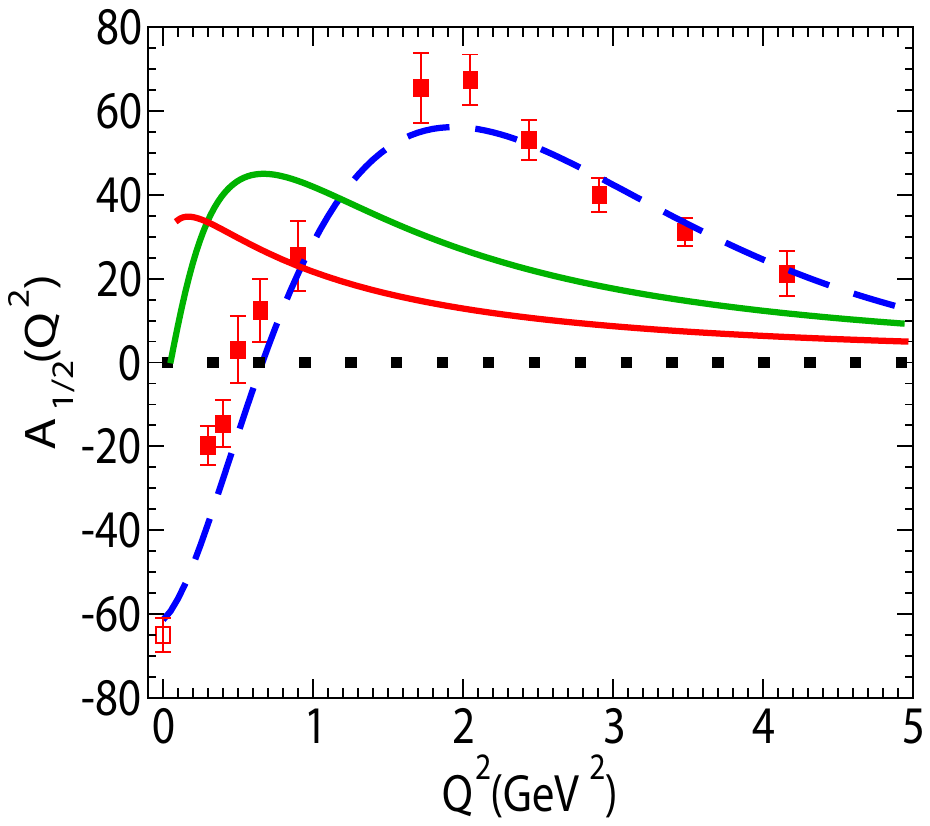}
        \includegraphics[width=6cm]{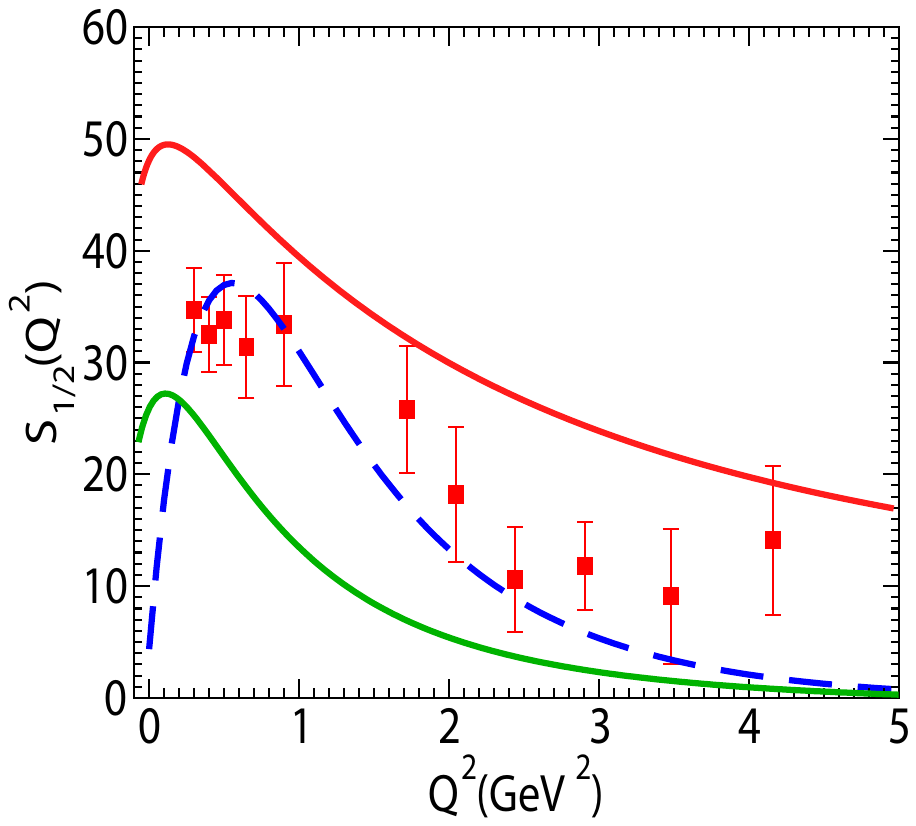}
    \end{center}
    \caption{$N+\gamma^{\ast}\longrightarrow N^{*}(1710)$ transition  $A_{1/2}(Q^2)$ and  $S_{1/2}(Q^2)$ helicity amplitudes in units of  $10^{-3}$ $ GeV^{-1/2}$ (red lines) is compared with CLAS experimental data (squares with error bars) [46], MAID fit (dashed lines) [40] and nonrelativistic quark model (green lines) [26].}
    \label{f2MB}
\end{figure}

\begin{figure}[!h]
    \begin{center}
        \includegraphics[width=6cm]{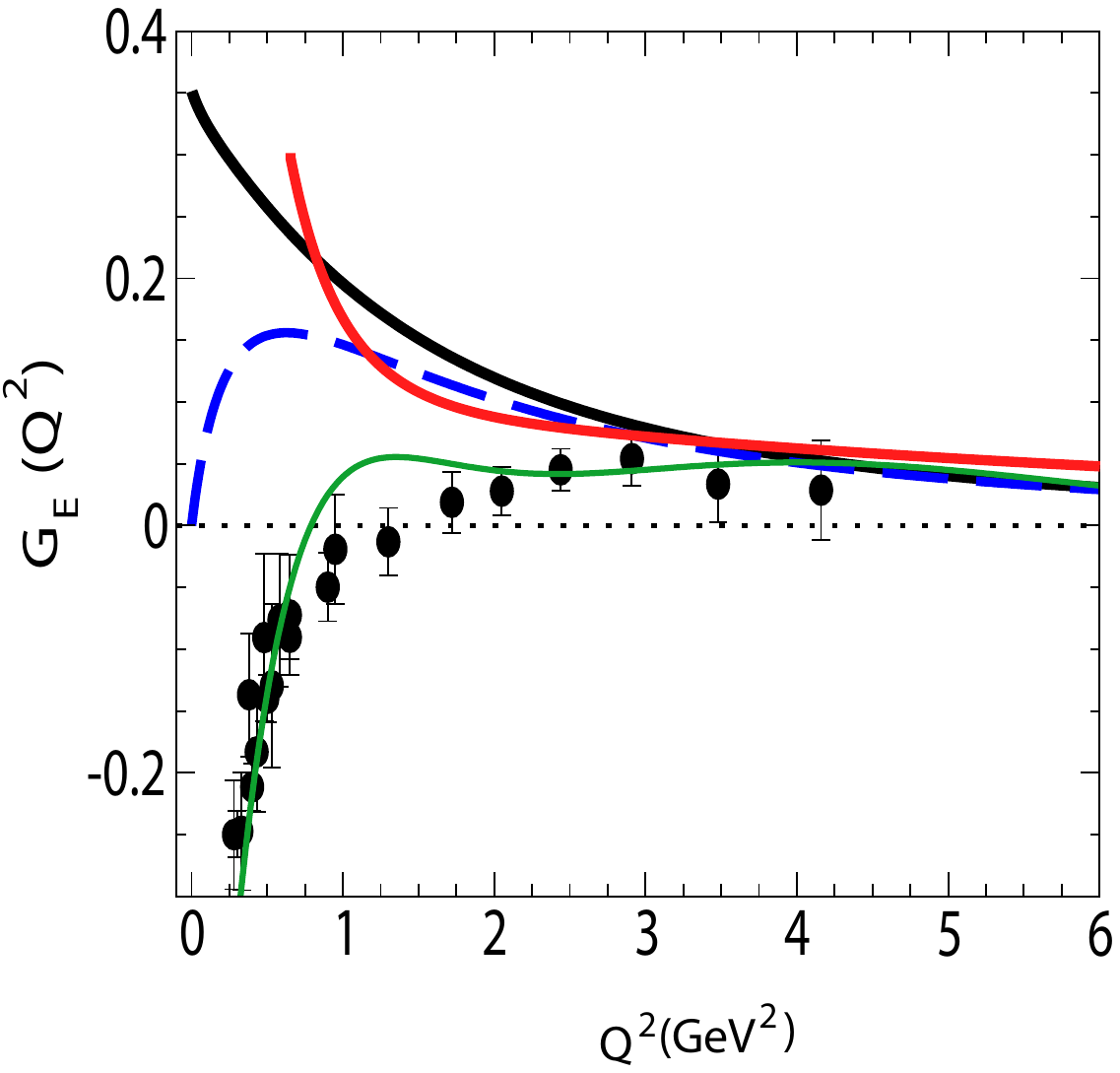}
        \includegraphics[width=6cm]{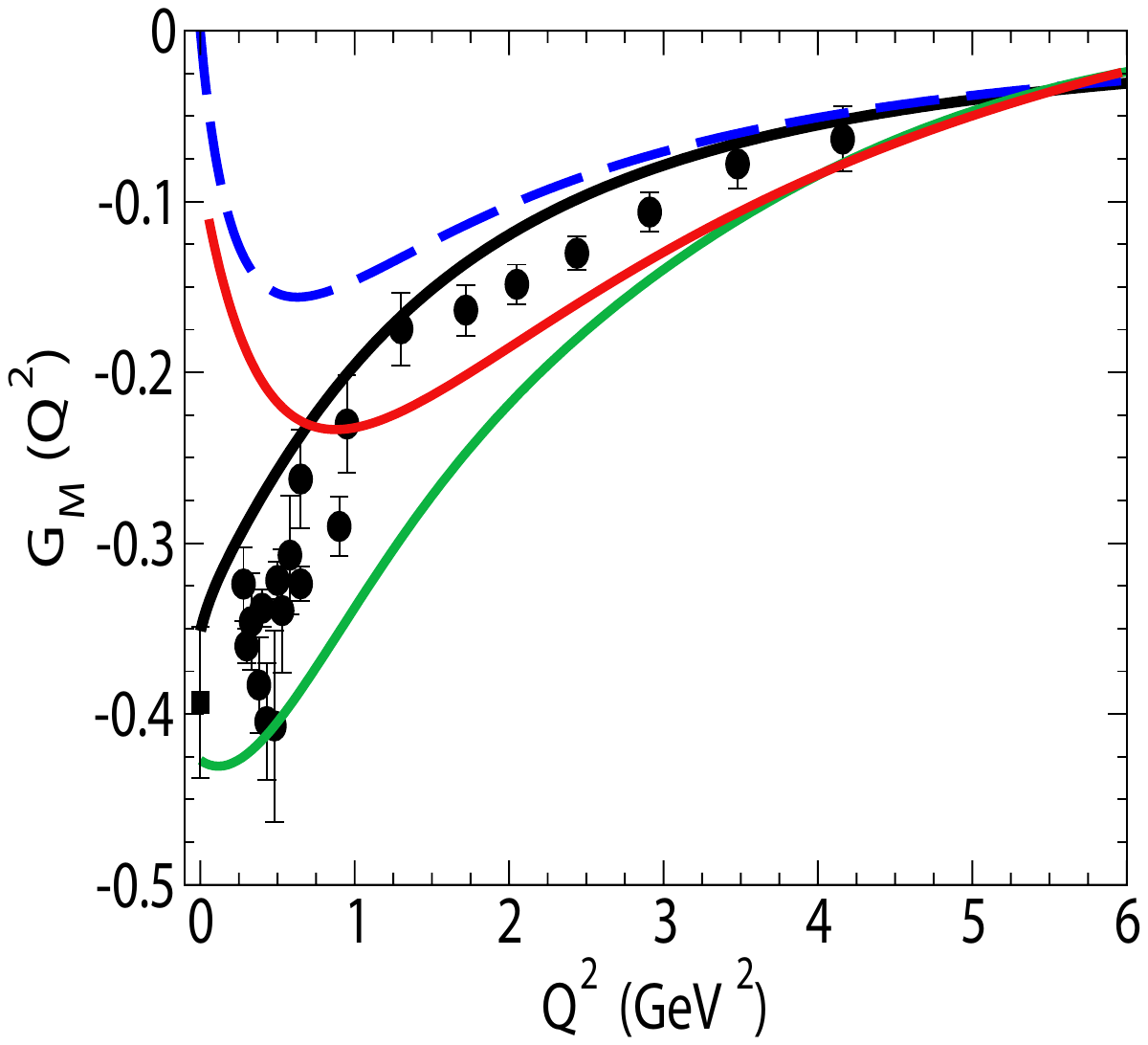}
    \end{center}
    \caption{The hard-wall model  results for the $G_{E}(Q^{2})$ and $G_{M}(Q^{2})$ form factors for the $N+\gamma^{\ast}\longrightarrow N^{*}(1535)$ with negative parity state (red line). Data from PDG [50] (full squares) and CLAS [10, 46, 51] (full circles). The thin solid line represent the fit to the MAID data [48] and the semirelativistic approximation (green thick solid line) [19].}
\end{figure}

\begin{figure}[!h]
    \begin{center}
        \includegraphics[width=6cm]{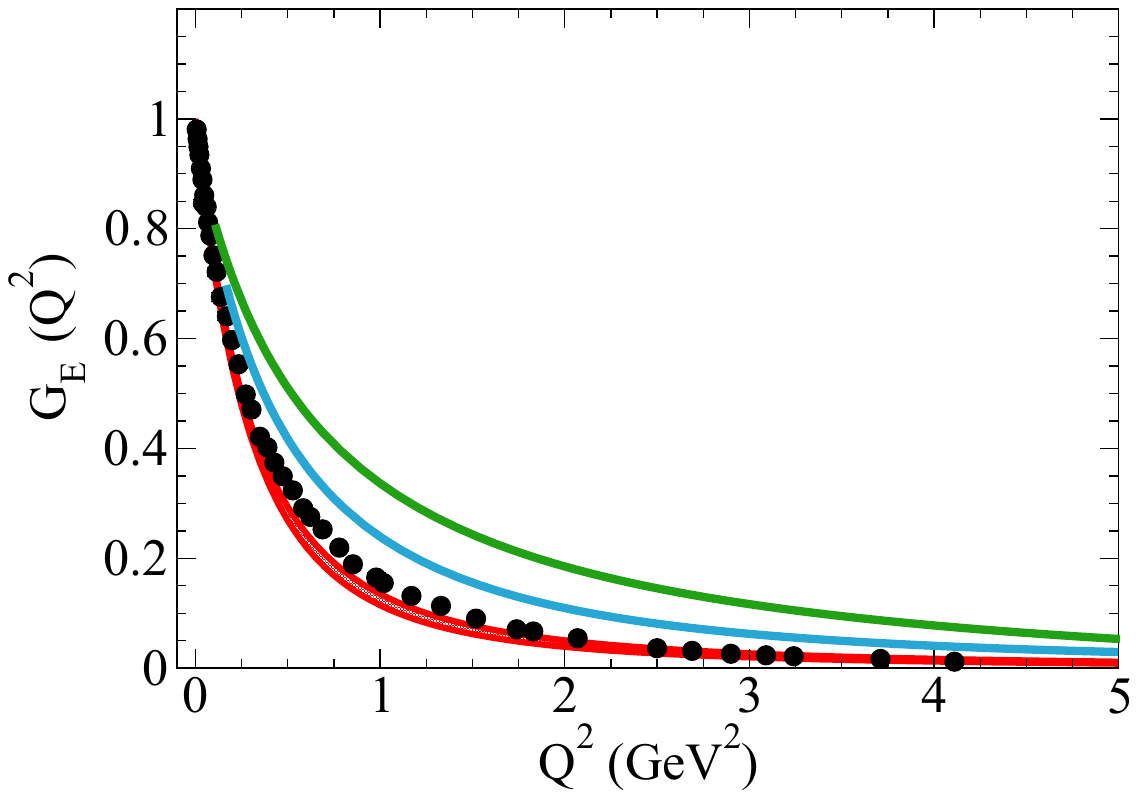}
        \includegraphics[width=6cm]{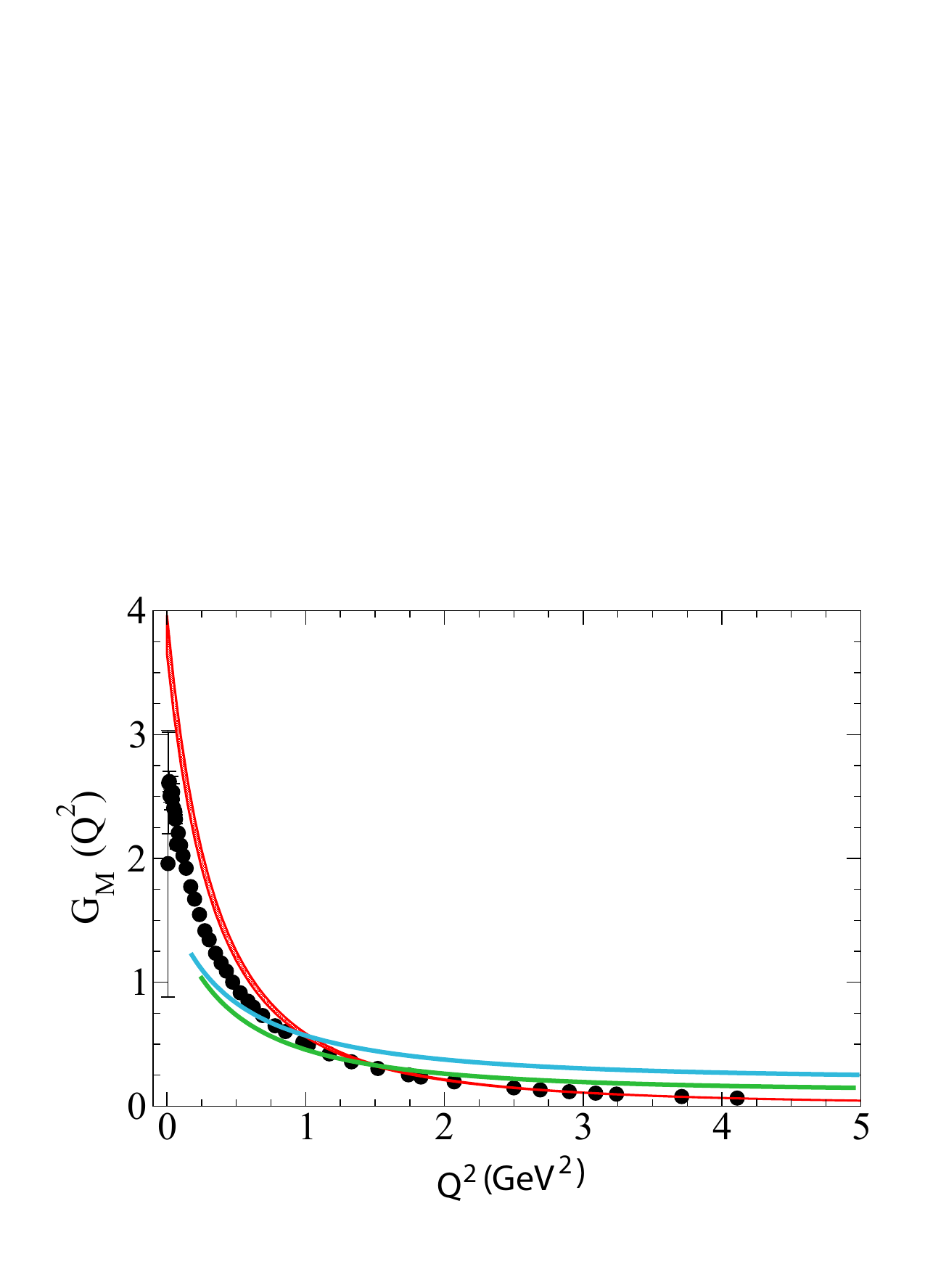}
    \end{center}
    \caption{The hard-wall model  results for the $G_{E}(Q^{2})$ and $G_{M}(Q^{2})$ form factors for the $N+\gamma^{\ast}\longrightarrow R(1440)$ (blue line) and $N+\gamma^{\ast}\longrightarrow N^{*}(1710)$ (green line) transitions. Data from Refs.[52-57]  and light front holography (red line) [17] }
\end{figure}

\begin{table}
\caption{Electromagnetic properties of the Roper nucleons}	
\begin{center}
\begin{tabular}{|c|c|c|c|c|c|c|}
  \hline
  Quantity & $ R (1440) $ & $ N^{*} (1535)$ & $ N^{*} (1710)$ & Soft-wall model  [21]  & Data  [9] & Experimental Data \\
  [0.5ex]
  \hline\hline
  $ r_{E}^{p}  (fm)$ & 0.8207& 0.7722 & 0.6640 & 0.840 & 0.8768 ± 0.0069   & 0.831 [58]   \\
  \hline
  $ r_{M}^{p}  (fm)$ &  0.687 & 0.638& 1.109 &0.785& 0.777 ± 0.013 ± 0.010 & 0.831  [59] \\
  \hline
  $\left \langle r_{E}^{2} \right \rangle^{n}  (fm^{2})$ & -0.566 & -0.312 & - 0.132  & -0.117 &-0.1161 ± 0.0022  & -0.111  [60]  \\
   \hline
   $ r_{M}^{n}  (fm)$   & 0.8207  &  0.7722 & 0.6640 & 0.792 & $0.862^{+0.009}_{-0.008}$ & 0.864  [61]  \\
  \hline
\end{tabular}
\end{center}
\end{table}

\end{document}